# A novel multi-party semiquantum private comparison protocol of size relationship with $d$-dimensional single-particle states

Tian-Yu Ye*, Jiang-Yuan Lian

College of Information & Electronic Engineering, Zhejiang Gongshang University, Hangzhou 310018, P.R.China

E-mail: happyyty@aliyun.com (T.Y. Ye)

**Abstract:** By using $d$-level single-particle states, the first multi-party semiquantum private comparison (MSQPC) protocol which can judge the size relationship of private inputs from more than two classical users within one execution of protocol is put forward. This protocol requires the help of one quantum third party (TP) and one classical TP, both of whom are allowed to misbehave on their own but cannot conspire with anyone else. Neither quantum entanglement swapping nor unitary operations are necessary for implementing this protocol. TPs are only required to perform $d$-dimensional single-particle measurements. The correctness analysis validates the accuracy of the compared results. The security analysis verifies that both the outside attacks and the participant attacks can be resisted.

## 1 Introduction

It is well known that quantum mechanics is one of the greatest scientific discoveries up to now. A novel kind of cryptography, which is called quantum cryptography, was put forward by combining quantum mechanics and classical cryptography in the year of 1984 [1]. In 1982, Yao [2] proposed the famous millionaire problem, which aims to determine who is richer on the basis that the wealth of two millionaires is not leaked out. The millionaire problem is essentially a problem of classical privacy comparison, whose security is based on the computational complexity of solving the corresponding mathematical problem. Later, in 2009, Yang and Wen [3] proposed the novel concept of quantum private comparison (QPC) by absorbing quantum mechanics into classical privacy comparison. Hereafter, a variety of QPC protocols [4-20] were proposed one after another. According to the function, there are two different types of QPC protocols, i.e., QPC of size relationship [4-10] and QPC of equality [3,11-20]. QPC of size relationship can determine the size relationship (i.e., greater than, smaller than and equal to) for the private inputs from different users, but QPC of equality just judges whether the private inputs from different users are identical or not. To some extent, QPC of size relationship may have wider applications than QPC of equality in reality.

In practice, not all users have the ability to acquire all kinds of quantum equipment. To get over this problem, Boyer *et al.* [21] proposed the innovative concept of semiquantum. In a semiquantum scheme, partial users are exempt from the preparation and measurement of quantum superposition states and quantum entangled states. Later, Ye *et al.* [22,23] employed single photons in two degrees of freedom to design two novel semiquantum key distribution (SQKD) protocols. In 2016, the first semiquantum private comparison (SQPC) scheme [24] was suggested by introducing the concept of semiquantum into QPC. Same to QPC, SQPC can be also classified into two different types: SQPC of equality [24-30] and SQPC of size relationship [31-35]. With respect to SQPC of size relationship, the ones in Refs.[32,33], Refs.[31,34] and Ref.[35] are based on $d$ - dimensional single-particle states, $d$ - dimensional Bell states and $d$ -dimensional GHZ states, respectively. Obviously, each of the SQPC protocols in Refs.[31-35] is only suitable for two classical users. At present, there is no SQPC protocol which can judge the size relationship for private inputs from more than two classical users within one execution of protocol.

Based on the above analysis, in this paper, we adopt $d$ -dimensional single-particle states to propose the first multi-party semiquantum private comparison (MSQPC) protocol which can judge the size relationship of private inputs from more than two classical users within one execution of protocol. One quantum third party (TP) and one classical TP, both of whom are allowed to misbehave on their own but cannot conspire with anyone else, help accomplish the comparison task. This protocol requires neither quantum entanglement swapping nor unitary operations. This protocol only needs TPs to perform $d$ -dimensional single-particle measurements.

## 2 Protocol description

In a $d$ -dimensional quantum system, the $Z$ -basis and the $X$ -basis can be described as

$$T_1 = \{|0\rangle, |1\rangle, \ldots, |d-1\rangle\} \tag{1}$$

and

$$T_2 = \{F|0\rangle, F|1\rangle, \ldots, F|d-1\rangle\}, \tag{2}$$

respectively. Here, $F$ is the $d$ - dimensional discrete quantum Fourier transform, and $F|t\rangle = \frac{1}{\sqrt{d}}\sum_{\delta=0}^{d-1} e^{\frac{2\pi i\delta t}{d}}|\delta\rangle$, $t = 0,1,\ldots,d-1$. $T_1$ and $T_2$ constitute two common conjugate bases.

Assume that there are $N$ classical users, $P_1, P_2, \ldots, P_N$, where $P_n$ possesses a $L$ -length private integer sequence $p_n = \{p_n^1, p_n^2, \ldots, p_n^L\}$. Here, $p_n^i \in \{0,1,\ldots,h\}$, $h = \frac{d-1}{2}$, $n = 1,2,\ldots,N$ and $i = 1,2,\ldots,L$. Besides, $P_1, P_2, \ldots, P_N$ pre-share a secret key sequence $K = \{k_1, k_2, \ldots, k_L\}$ through a secure mediated SQKD protocol [36], where $k_i \in \{0,1,\ldots,d-1\}$ and $i=1,2,\ldots,L$. In the proposed MSQPC protocol, $TP_1$ is the quantum TP who has full quantum capabilities while $TP_2$ is the classical TP who merely possesses limited quantum abilities. By virtue of Ref.[37], $TP_1$ and $TP_2$ are permitted to launch all types of attacks according to their own will but cannot collude with anyone else. We describe the proposed MSQPC protocol in detail as follows.

Step 1: $TP_1$ prepares $N$ singe-particle state sequences whose particles are randomly chosen from two sets $T_1$ and $T_2$. These $N$ singe-particle state sequences are represented by $S_1, S_2, \ldots, S_N$, where $S_n = \{q_n^1, q_n^2, \ldots, q_n^{16L}\}$ and $n=1,2,\ldots,N$. Then, $TP_1$ sends $S_n$ to $P_n$ via a quantum channel. Note that except the first particle, $TP_1$ sends out the next particle of $S_n$ to $P_n$ only after receiving the previous one from $TP_2$.

Step 2: $P_n$ produces a random binary sequence $r_n$, where $r_n = \{r_n^1, r_n^2, \ldots, r_n^{16L}\}$, $r_n^l \in \{0,1\}$, $n=1,2,\ldots,N$ and $l=1,2,\ldots,16L$. After receiving the $l$ th particle of $S_n$, $P_n$ enters into the REFLRCT mode or the MEASURE mode according to $r_n^l$. To be specific, when $r_n^l = 0$, $P_n$ selects the REFLECT mode; otherwise, $P_n$ chooses the MEASURE mode. Here, the RFFLECT mode replies to reflecting the received particle back to the sender without any interference, while the MEASURE mode means to measuring the received particle in the $T_1$ basis, preparing the same quantum state as found and sending it back to the sender. Note that $P_n$ needs to record her measurement results when entering into the MEASURE mode. The new sequence after $P_n$ performs her operations on $S_n$ is represented by $S_n^{'}$, where $S_n^{'} = \{q_n^{1'}, q_n^{2'}, \ldots, q_n^{16L'}\}$. Finally, $P_n$ sends $S_n^{'}$ to $TP_2$ via a quantum channel.

Step 3: $TP_2$ produces a random binary sequence $v_n$, where $v_n = \{v_n^1, v_n^2, \ldots, v_n^{16L}\}$, $v_n^l \in \{0,1\}$, $n=1,2,\ldots,N$ and $l=1,2,\ldots,16L$. After receiving the $l$ th particle of $S_n^{'}$, $TP_2$ enters into the REFLRCT mode or the MEASURE mode according to $v_n^l$. Concretely speaking, when $v_n^l = 0$, $TP_2$ selects the REFLECT mode; otherwise, $P_n$ chooses the MEASURE mode. Note that when choosing the MEASURE mode, $TP_2$ should record her measurement results. The new sequence after $TP_2$ 's operations on $S_n^{'}$ is denoted as $S_n^{''}$, where $S_n^{''} = \{q_n^{1''}, q_n^{2''}, \ldots, q_n^{16L''}\}$. Finally, $TP_2$ sends $S_n^{''}$ to $TP_1$ via a quantum channel.

Step 4: $TP_1$ publishes the positions of particles prepared within the set $T_2$ in Step 1. In the meanwhile, $P_n$ and $TP_2$ announce $r_n$ and $v_n$, respectively, where $n=1,2,\ldots,N$. Based on the announced information, $TP_1$ executes the corresponding operations as listed in Table 1.

Case 1: In this case, the initial particles are prepared by $TP_1$ within the $T_1$ basis in Step 1; both $P_n$ and $TP_2$ have selected the REFLECT mode; and $TP_1$ measures the corresponding particles in her hand with the $T_1$ basis. By comparing her measurements with the corresponding initial prepared states, $TP_1$ can judge whether there is an eavesdropper or not. If there is no eavesdropper, this protocol will be proceeded;

Case 2: In this case, the initial particles are prepared by $TP_1$ within the $T_2$ basis in Step 1; both $P_n$ and $TP_2$ have selected the REFLECT mode; and $TP_1$ measures the corresponding particles in her hand with the $T_2$ basis. By comparing her measurements with the corresponding initial prepared states, $TP_1$ can judge whether there is an eavesdropper or not. If there is no eavesdropper, this protocol will be proceeded;

Case 3: In this case, the initial particles are prepared by $TP_1$ within the $T_1$ basis in Step 1; $P_n$ and $TP_2$ have chosen the MEASURE mode and the REFLECT mode, respectively; and $TP_1$ measures the corresponding particles in her hand with the $T_1$ basis. $P_n$ needs to tell $TP_1$ the states of the newly generated particles. $TP_1$ compares her measurement results with the states of the newly generated particles from $P_n$ and the corresponding initial prepared states. If there is no eavesdropper, this protocol will be proceeded;

Case 4: In this case, the initial particles are prepared by $TP_1$ within the $T_1$ basis in Step 1; $P_n$ and $TP_2$ have chosen the REFLECT mode and the MEASURE mode, respectively; and $TP_1$ measures the corresponding particles in her hand with the $T_1$ basis. $TP_2$ needs to tell $TP_1$ the states of the newly generated particles. $TP_1$ compares her measurement results with the states of the newly generated particles from $TP_2$ and the corresponding initial prepared states. If there is no eavesdropper, this protocol will be proceeded;

Case 5, Case 6 and Case 7: In these three Cases, the initial particles are prepared by $TP_1$ within the $T_2$ basis in Step 1; at least one party from $P_n$ and $TP_2$ has chosen the MEASURE mode; and $TP_1$ takes no action. Note that these three Cases are ignored;

Case 8: In this case, the initial particles are prepared by $TP_1$ within the $T_1$ basis in Step 1; both $P_n$ and $TP_2$ have chosen the MEASURE mode; and $TP_1$ measures the corresponding particles in her hand with the $T_1$ basis. If the number of the corresponding particles in this Case on the site of $TP_1$ is less than $2L$, the protocol will be suspended.

Step 5: $TP_1$ randomly picks out $L$ particles from the ones belonging to Case 8 in her hand, and publishes their positions. Then, $P_n$ and $TP_2$ announce their measurement results on these chosen positions, respectively. Afterward, $TP_1$ checks the error rate for these chosen particles by comparing her measurement results with $P_n$ and $TP_2$'s measurement results and the corresponding initial prepared states. If the error rate is zero, they will continue the protocol.

Step 6: $P_n$, $TP_1$ and $TP_2$ use the remaining $L$ particles in Case 8 for private comparison. Note that all of $P_n$, $TP_1$ and $TP_2$'s measurement results on the particles in Case 8 are same. $P_n$, $TP_1$ and $TP_2$'s measurement results on the remaining $L$ particles in Case 8 are denoted as $m_n = \{|m_n^1\rangle, |m_n^2\rangle, \ldots, |m_n^L\rangle\}$, where $m_n^i \in \{0, 1, \ldots, d-1\}$, $n = 1, 2, \ldots, N$ and $i = 1, 2, \ldots, L$. $P_n$ calculates

$$c_n^i = m_n^i \oplus k_i \oplus p_n^i, \tag{3}$$

where the symbol $\oplus$ denotes the modulo $d$ addition. Finally, $P_n$ sends $c_n$ to $TP_1$ via an authenticated classical channel, where $c_n = \{c_n^1, c_n^2, \ldots, c_n^L\}$.

Table 1 $TP_1$'s actions under different Cases

| Case | $TP_1$'s preparation basis | $r_n^l$ | $v_n^l$ | $TP_1$'s action |
|---|---|---|---|---|
| Case 1 | The $T_1$ basis | 0 | 0 | Measuring $q_n^{l''}$ with the $T_1$ basis |
| Case 2 | The $T_2$ basis | 0 | 0 | Measuring $q_n^{l''}$ with the $T_2$ basis |
| Case 3 | The $T_1$ basis | 1 | 0 | Measuring $q_n^{l''}$ with the $T_1$ basis |
| Case 4 | The $T_1$ basis | 0 | 1 | Measuring $q_n^{l''}$ with the $T_1$ basis |
| Case 5 | The $T_2$ basis | 0 | 1 | Ignored |
| Case 6 | The $T_2$ basis | 1 | 0 | Ignored |
| Case 7 | The $T_2$ basis | 1 | 1 | Ignored |
| Case 8 | The $T_1$ basis | 1 | 1 | Measuring $q_n^{l''}$ with the $T_1$ basis |

Step 7: After receiving $c_n$, for $n = 1, 2, \ldots, N$ and $i = 1, 2, \ldots, L$, $TP_1$ calculates

$$f_n^i = c_n^i \ominus m_n^i. \tag{4}$$

Then, $TP_1$ computes

$$R_{nn'}^{i} = f_n^i \ominus f_{n'}^{i}, \tag{5}$$

where $n' = 1, 2, \ldots, N$ and $n' \neq n$. $TP_1$ makes

$$y\left(R_{nn'}^{i}\right) = \begin{cases} -1, & if \quad h < R_{nn'}^{i} \leq 2h; \\ 0, & if \quad R_{nn'}^{i} = 0; \\ 1, & if \quad 0 < R_{nn'}^{i} \leq h. \end{cases} \tag{6}$$

Here, $y\left(R_{nn'}^{i}\right) = -1$ means $p_n^i < p_{n'}^{i}$; $y\left(R_{nn'}^{i}\right) = 0$ means $p_n^i = p_{n'}^{i}$; $y\left(R_{nn'}^{i}\right) = 1$ means $p_n^i > p_{n'}^{i}$. Finally, $TP_1$ publishes the final comparison results to $P_1, P_2, \ldots, P_N$.

To be more clearly, the flow chart of the proposed MSQPC protocol is drawn in Fig.1 after the processes of security check are ignored.

## 3 Correctness analysis

### 3.1 Output correctness

After inserting Eq.(3) and Eq.(4) into Eq.(5), we can obtain

$$\begin{aligned} R_{nn'}^{i} &= f_n^i \ominus f_{n'}^{i} \\ &= \left(c_n^i \ominus m_n^i\right) \ominus \left(c_{n'}^{i} \ominus m_{n'}^{i}\right) \\ &= \left(m_n^i \oplus k_i \oplus p_n^i \ominus m_n^i\right) \ominus \left(m_{n'}^{i} \oplus k_i \oplus p_{n'}^{i} \ominus m_{n'}^{i}\right) \\ &= \left(k_i \oplus p_n^i\right) \ominus \left(k_i \oplus p_{n'}^{i}\right) \\ &= p_n^i \ominus p_{n'}^{i}. \end{aligned} \tag{7}$$

Here, $n, n' = 1, 2, \ldots, N$, $n' \neq n$ and $i = 1, 2, \ldots, L$. Since $p_n^i, p_{n'}^{i} \in \{0, 1, 2, \ldots, h\}$ and $h = \frac{d-1}{2}$, according to Eq.(6) and Eq.(7), the following points stand: when $y\left(R_{nn'}^{i}\right) = -1$, it has $h < p_n^i \ominus p_{n'}^{i} \leq 2h$, which implies $p_n^i < p_{n'}^{i}$; when $y\left(R_{nn'}^{i}\right) = 0$, it has $p_n^i \ominus p_{n'}^{i} = 0$, which implies $p_n^i = p_{n'}^{i}$; and when $y\left(R_{nn'}^{i}\right) = 1$, it has $0 < p_n^i \ominus p_{n'}^{i} \leq h$, which implies $p_n^i > p_{n'}^{i}$. It can be concluded that the comparison result of the proposed MSQPC protocol is correct.

### 3.2 Examples

To further prove the correctness of the proposed MSQPC protocol, we give a specific example.

Assume that the dimension of quantum system is $d=19$; there are four classical users, $P_1,P_2,P_3,P_4$; the first private integers of $P_1,P_2,P_3,P_4$ are $p_1^1=5$, $p_2^1=3$, $p_3^1=5$ and $p_4^1=6$, respectively; the first private key integer pre-shared among $P_1,P_2,P_3,P_4$ is $k_1=16$; the measurement results of $P_1,P_2,P_3,P_4$ on the first remaining particles in Case 8 are $\left|m_1^1\right\rangle=|7\rangle$, $\left|m_2^1\right\rangle=|2\rangle$, $\left|m_3^1\right\rangle=|9\rangle$ and $\left|m_4^1\right\rangle=|10\rangle$, respectively. According to Eq.(3), $P_1,P_2,P_3,P_4$ can obtain $c_1^1=7\oplus16\oplus5=9$, $c_2^1=2\oplus16\oplus3=2$, $c_3^1=9\oplus16\oplus5=11$ and $c_4^1=10\oplus16\oplus6=13$, respectively. Then, $P_1,P_2,P_3,P_4$ send $c_1^1,c_2^1,c_3^1,c_4^1$ to $TP_1$ via an authenticated classical channel, respectively. After receiving $c_1^1,c_2^1,c_3^1,c_4^1$, by virtue of Eq.(4), $TP_1$ can get $f_1^1=c_1^1\ominus m_1^1=9\ominus7=2$, $f_2^1=c_2^1\ominus m_2^1=2\ominus2=0$, $f_3^1=c_3^1\ominus m_3^1=11\ominus9=2$ and $f_4^1=c_4^1\ominus m_4^1=13\ominus10=3$. Afterward, through Eq.(5), $TP_1$ can obtain $R_{12}^1=f_1^1\ominus f_2^1=2\ominus0=2$, $R_{13}^1=f_1^1\ominus f_3^1=2\ominus2=0$, $R_{14}^1=f_1^1\ominus f_4^1=2\ominus3=18$, $R_{23}^1=f_2^1\ominus f_3^1=0\ominus2=17$, $R_{24}^1=f_2^1\ominus f_4^1=0\ominus3=16$ and $R_{34}^1=f_3^1\ominus f_4^1=2\ominus3=18$. According to Eq.(6), $TP_1$ makes $y\left(R_{12}^1\right)=1$, $y\left(R_{13}^1\right)=0$, $y\left(R_{14}^1\right)=-1$, $y\left(R_{23}^1\right)=-1$, $y\left(R_{24}^1\right)=-1$ and $y\left(R_{34}^1\right)=-1$, which imply $p_1^1>p_2^1$, $p_1^1=p_3^1$, $p_1^1<p_4^1$, $p_2^1<p_3^1$, $p_2^1<p_4^1$ and $p_3^1<p_4^1$, respectively. In short, it has $p_2^1<p_1^1=p_3^1<p_4^1$.

## 4 Security analysis

### 4.1 Outside attacks

An outsider Eve may do her best to acquire $p_n$ ($n=1,2,\ldots,N$) by launching some famous attacks, e.g., the intercept-resend attack, the measure-resend attack and the entangle-measure attack.

(1) The intercept-resend attack

There are three types of intercept-resend attacks according to the process of the proposed protocol. We will do the detailed analysis for them one by one in the following.

Firstly, Eve intercepts the particle of $S_n$ and sends the fake one she has already prepared in the $T_1$ basis to $P_n$ in Step 1; after $P_n$ executes her operation, Eve intercepts the particle of $S_n^{'}$ and sends the original genuine one to $TP_2$ in Step 2. When $P_n$ has chosen the REFLECT mode, Eve's attack cannot be detected, no matter what the original genuine particle is and which mode $TP_2$ has chosen. Consider the situation that $P_n$ has chosen the MEASURE mode: if the original genuine particle is in the $T_2$ basis, as ignored according to Table 1, Eve's attack will not be detected in Step 4; if the original genuine particle is in the $T_1$ basis, Eve's attack will be discovered with the probability of $\frac{d-1}{d}$ in Step 4 and the probability of $\frac{d-1}{2d}$ in Step 5 when $TP_2$ has chosen the REFLECT mode and the MEASURE mode in Step 3, respectively.

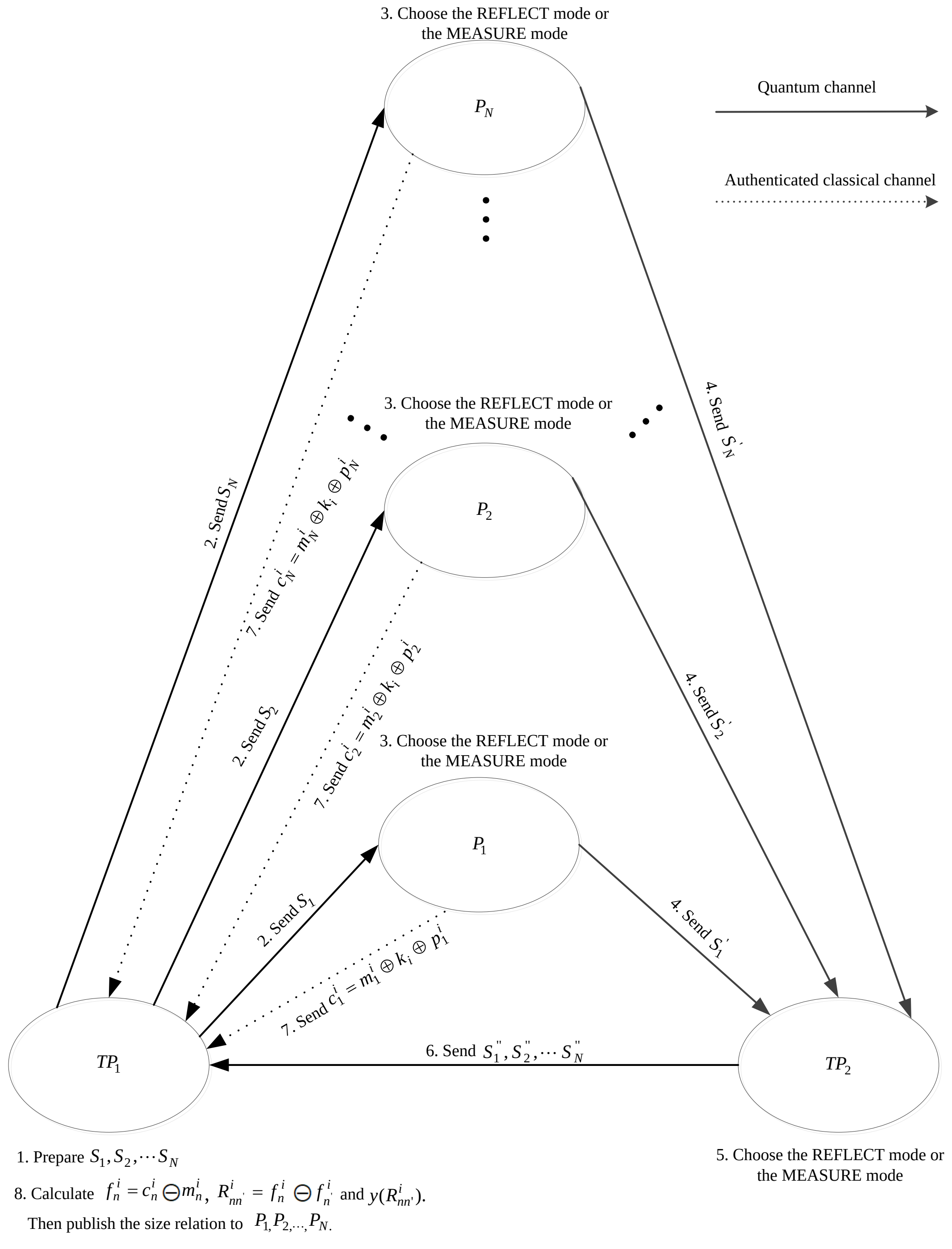

Fig.1 The flow chart of the proposed MSQPC protocol

Secondly, Eve intercepts the particle of $S_n$ and sends the fake one she has generated in advance in the $T_1$ basis to $P_n$ in Step 1; then in Step 3, Eve intercepts the particle of $S_n^{''}$ and sends the original genuine one to $TP_1$. Considering that the original genuine particle is in the $T_1$ basis, if $P_n$ and $TP_2$ have chosen the REFLECT mode and the MEASURE mode, respectively, the existence of Eve will be detected with the probability of $\frac{d-1}{d}$ in Step 4; if both $P_n$ and $TP_2$ have chosen the MEASURE mode, the existence of Eve will be detected with the probability of $\frac{d-1}{2d}$ in Step 5; if $P_n$ and $TP_2$ have chosen the MEASURE mode and the REFLECT mode, respectively, the existence of Eve will be detected with the probability of $\frac{d-1}{d}$ in Step 4; if both $P_n$ and $TP_2$ have chosen the REFLECT mode, the existence of Eve will not be detected in Step 4. Considering that the original genuine particle is in the $T_2$ basis, no matter what mode $P_n$ and $TP_2$ have chosen, the existence of Eve cannot be detected in Step 4.

Thirdly, Eve intercepts the particle of $S_n^{'}$ and sends the fake one she has generated in advance in the $T_1$ basis to $TP_2$ in Step 2; after $TP_2$ applies her operation on the fake one, Eve intercepts the particle sent from $TP_2$ and sends the genuine one to $TP_1$ in Step 3. Considering that the original genuine particle is in the $T_1$ basis, if both $P_n$ and $TP_2$ have chosen the MEASURE mode, the presence of Eve will be detected with the probability of $\frac{d-1}{2d}$ in Step 5; if $P_n$ and $TP_2$ have chosen the REFLECT mode and the MEASURE mode, respectively, the presence of Eve will be detected with the probability of $\frac{d-1}{d}$ in Step 4; if $P_n$ and $TP_2$ have chosen the MEASURE mode and the REFLECT mode, respectively, the presence of Eve will be detected with the probability of 0 in Step 4; if both $P_n$ and $TP_2$ have chosen the REFLECT mode, the presence of Eve will be detected with the probability of 0 in Step 4. Considering that the original genuine particle is in the $T_2$ basis, no matter what mode $P_n$ and $TP_2$ have chosen, the existence of Eve cannot be detected in Step 4.

(2) The measure-resend attack

In the following, we analyze three kinds of measure-resend attacks.

Eve intercepts the particle of $S_n / S_n^{'} / S_n^{''}$, measures it with the $T_1$ basis and sends the resulted state to $P_n / TP_2 / TP_1$. If the original particle is in the $T_1$ basis, this attack cannot be discovered, no matter what mode $P_n$ and $TP_2$ have chosen. Considering that the original particle is in the $T_2$ basis, if at least one of $P_n$ and $TP_2$ has selected the MEASURE mode, Eve's attack will not be detected either; if both $P_n$ and $TP_2$ have chosen the REFLECT mode, Eve's attack will be detected in Step 4, as the state of original particle was destroyed by Eve's measurement.

(3) The entangle-measure attack

Eve may launch her entangle-measure attack, as shown in Fig.2, by using two unitaries, $U_E$ and $U_F$, where $U_E$ and $U_F$ share a common probe space with the state $|E\rangle$. Here, Eve applies $U_E$ on the particle from $TP_1$ to $P_n$ and performs $U_F$ on the particle from $P_n$ to $TP_2$. As depicted in Ref.[21], Eve is allowed by the shared probe to perform the attack on the particle which are transmitted back according to the knowledge acquired by $U_E$.

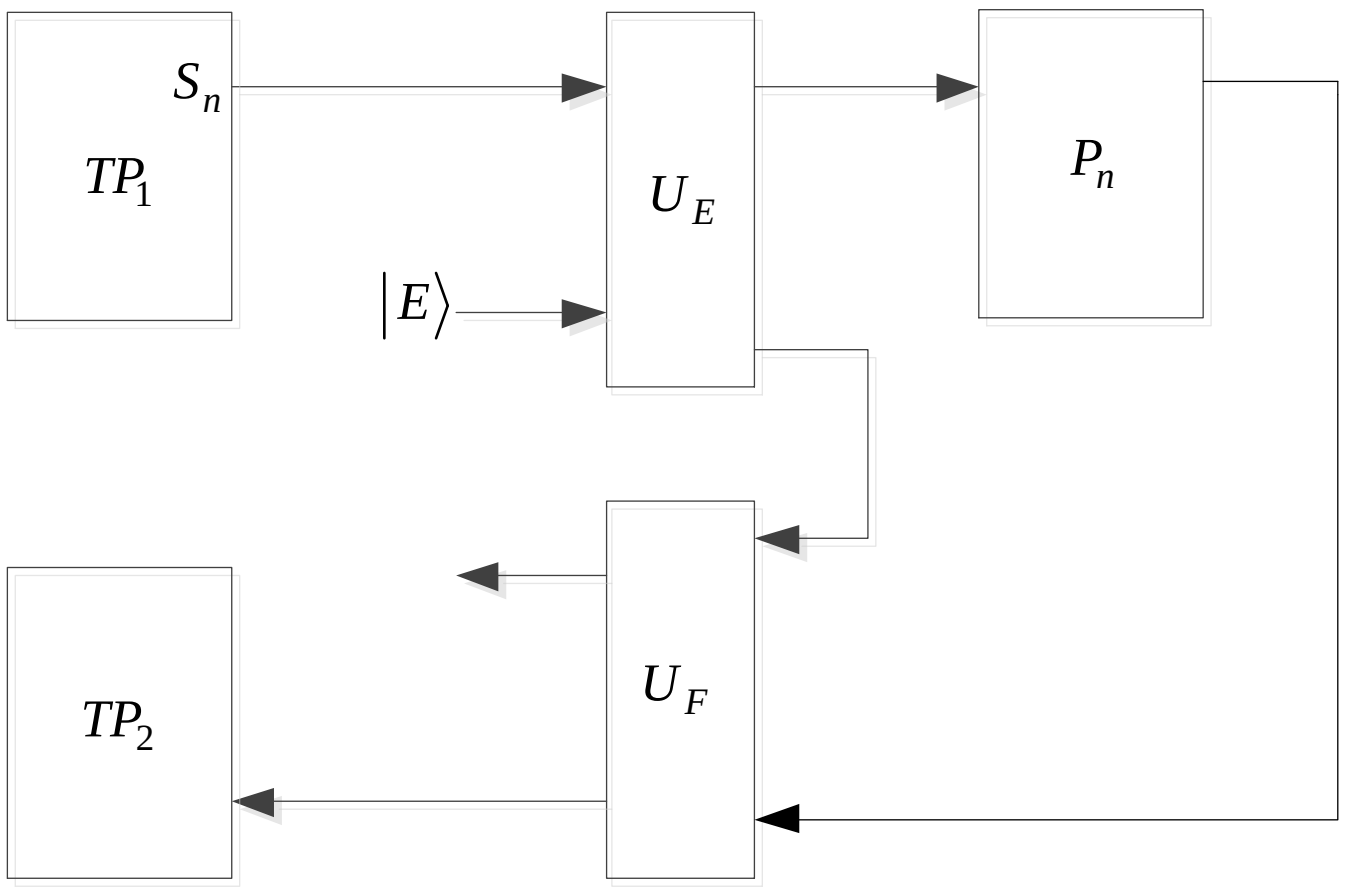

Fig.2 Eve's entangle-measure attack with $U_E$ and $U_F$

**Theorem 1.** *Suppose that Eve applies $U_E$ on the particle from $TP_1$ to $P_n$ and performs $U_F$ on the particle from $P_n$ to $TP_2$. For introducing no error in Step 4 and Step 5, the final state of Eve's probe should be independent of not only the operation of $P_n$ but also the measurement results of $P_n$ and $TP_2$. As a result, Eve has no access to $m_n$.*

**Proof.** For simplicity, we use $|t\rangle$ and $|J_t\rangle$ to denote the particle in the sets $T_1$ and $T_2$, respectively, where $|J_t\rangle = F|t\rangle = \frac{1}{\sqrt{d}}\sum_{\delta=0}^{d-1} e^{\frac{2\pi i\delta t}{d}}|\delta\rangle$ and $t = 0,1,\ldots,d-1$.

(1) Consider the situation that the particle of $S_n$ is prepared in the $T_1$ basis. When $TP_1$ sends out the particle of $S_n$, Eve performs $U_E$ on it, thus we can get [38]

$$U_E\left(|t\rangle|E\rangle\right)=\sum_{t'=0}^{d-1}\gamma_{tt'}\left|t'\right\rangle\left|e_{tt'}\right\rangle, \tag{8}$$

where $\left|e_{tt'}\right\rangle$ ($t,t'=0,1,\ldots,d-1$) are the probe states of Eve decided by $U_E$, and

$$\sum_{t'=0}^{d-1}\left|\gamma_{tt'}\right|^2=1. \tag{9}$$

When $P_n$ performs the MEASURE operation, the global composite system is collapsed into $\gamma_{tt'}\left|t'\right\rangle\left|e_{tt'}\right\rangle$. In order to avoid being discovered during the security checks of Case 3 and Case 8, after Eve performs $U_F$, the global state of the composite system should satisfy

$$U_F\left(\gamma_{tt'}\left|t'\right\rangle\left|e_{tt'}\right\rangle\right)=\begin{cases}\gamma_{tt}|t\rangle|F_{tt}\rangle, & if \quad t=t'; \\ 0, & if \quad t\neq t'.\end{cases}, \tag{10}$$

which implies

$$\gamma_{tt'}=\begin{cases}\gamma_{tt}, & if \quad t=t'; \\ 0, & if \quad t\neq t'.\end{cases}. \tag{11}$$

When $P_n$ performs the REFLECT operation, according to Eq.(8) and Eq.(10), the global composite system after Eve performs $U_F$ is turned into

$$U_F\left[U_E\left(|t\rangle|E\rangle\right)\right]=U_F\left(\sum_{t'=0}^{d-1}\gamma_{tt'}\left|t'\right\rangle\left|e_{tt'}\right\rangle\right)=\sum_{t'=0}^{d-1}U_F\left(\gamma_{tt'}\left|t'\right\rangle\left|e_{tt'}\right\rangle\right)=\gamma_{tt}|t\rangle|F_{tt}\rangle \tag{12}$$

for $t=0,1,\ldots,d-1$. In order to avoid being discovered during the security checks of Case 1 and Case 4, Eve cannot change the state of the original particle of $S_n$. According to Eq.(12), this requirement is automatically satisfied.

(2) Consider the situation that the particle of $S_n$ is prepared in the $T_2$ basis. After Eve performs $U_E$, the global composite system is evolved into

$$\begin{aligned}U_E\left(|J_t\rangle|E\rangle\right)&=U_E\left[\left(\frac{1}{\sqrt{d}}\sum_{\delta=0}^{d-1}e^{\frac{2\pi i\delta t}{d}}|\delta\rangle\right)|E\rangle\right]\\&=\frac{1}{\sqrt{d}}\sum_{\delta=0}^{d-1}e^{\frac{2\pi i\delta t}{d}}U_E\left(|\delta\rangle|E\rangle\right).\end{aligned} \tag{13}$$

When $P_n$ chooses the REFLECT operation, the global composite system after Eve performs $U_F$ is turned into

$$U_F\left[U_E\left(\left|J_t\right\rangle\left|E\right\rangle\right)\right]=\frac{1}{\sqrt{d}}\sum_{\delta=0}^{d-1}e^{\frac{2\pi i\delta t}{d}}U_F\left[U_E\left(\left|\delta\right\rangle\left|E\right\rangle\right)\right]. \tag{14}$$

Inserting Eq.(12) into Eq.(14) generates

$$U_F\left[U_E\left(\left|J_t\right\rangle\left|E\right\rangle\right)\right]=\frac{1}{\sqrt{d}}\sum_{\delta=0}^{d-1}e^{\frac{2\pi i\delta t}{d}}\gamma_{\delta\delta}\left|\delta\right\rangle\left|F_{\delta\delta}\right\rangle. \tag{15}$$

By virtue of the inverse quantum Fourier transform, we have

$$\left|\delta\right\rangle=\frac{1}{\sqrt{d}}\sum_{\alpha=0}^{d-1}e^{-\frac{2\pi i\alpha\delta}{d}}\left|J_\alpha\right\rangle, \tag{16}$$

where $\delta=0,1,\ldots,d-1$. Inserting Eq.(16) into Eq.(15) produces

$$\begin{aligned}U_F\left[U_E\left(\left|J_t\right\rangle\left|E\right\rangle\right)\right]&=\frac{1}{\sqrt{d}}\sum_{\delta=0}^{d-1}e^{\frac{2\pi i\delta t}{d}}\gamma_{\delta\delta}\left(\frac{1}{\sqrt{d}}\sum_{\alpha=0}^{d-1}e^{-\frac{2\pi i\alpha\delta}{d}}\left|J_\alpha\right\rangle\right)\left|F_{\delta\delta}\right\rangle\\&=\frac{1}{d}\sum_{\delta=0}^{d-1}\sum_{\alpha=0}^{d-1}e^{\frac{2\pi i\delta(t-\alpha)}{d}}\gamma_{\delta\delta}\left|J_\alpha\right\rangle\left|F_{\delta\delta}\right\rangle\\&=\frac{1}{d}\left(\left|J_0\right\rangle\sum_{\delta=0}^{d-1}e^{\frac{2\pi i\delta(t-0)}{d}}\gamma_{\delta\delta}\left|F_{\delta\delta}\right\rangle+\left|J_1\right\rangle\sum_{\delta=0}^{d-1}e^{\frac{2\pi i\delta(t-1)}{d}}\gamma_{\delta\delta}\left|F_{\delta\delta}\right\rangle+\ldots+\left|J_{d-1}\right\rangle\sum_{\delta=0}^{d-1}e^{\frac{2\pi i\delta[t-(d-1)]}{d}}\gamma_{\delta\delta}\left|F_{\delta\delta}\right\rangle\right).\end{aligned} \tag{17}$$

In order that Eve cannot be discovered during the security check of Case 2, it should meet

$$\sum_{\delta=0}^{d-1}e^{\frac{2\pi i\delta(t-\alpha)}{d}}\gamma_{\delta\delta}\left|F_{\delta\delta}\right\rangle=0 \tag{18}$$

for $t\neq\alpha$ and $t,\alpha=0,1,\ldots,d-1$. It is obvious that for $t\neq\alpha$, we have

$$\sum_{\delta=0}^{d-1}e^{\frac{2\pi i\delta(t-\alpha)}{d}}=0. \tag{19}$$

Thus, by virtue of Eq.(18) and Eq.(19), we have

$$\gamma_{00}\left|F_{00}\right\rangle=\gamma_{11}\left|F_{11}\right\rangle=\ldots=\gamma_{(d-1)(d-1)}\left|F_{(d-1)(d-1)}\right\rangle=\gamma\left|F\right\rangle. \tag{20}$$

(3) Applying Eq.(20) into Eq.(10) generates

$$U_F\left(\gamma_{tt'}\left|t'\right\rangle\left|e_{tt'}\right\rangle\right)=\begin{cases}\gamma\left|t\right\rangle\left|F\right\rangle, & if \quad t=t';\\ 0, & if \quad t\neq t'.\end{cases} \tag{21}$$

Applying Eq.(20) into Eq.(12) creates

$$U_F\left[U_E\left(\left|t\right\rangle\left|E\right\rangle\right)\right]=\gamma\left|t\right\rangle\left|F\right\rangle. \tag{22}$$

Applying Eq.(20) into Eq.(17) produces

$$U_F\left[U_E\left(\left|J_t\right\rangle\left|E\right\rangle\right)\right]=\gamma\left|J_t\right\rangle\left|F\right\rangle. \tag{23}$$

Based on Eq.(21), Eq.(22) and Eq.(23), when Eve applies $U_E$ on the particle from $TP_1$ to $P_n$ and performs $U_F$ on the particle from $P_n$ to $TP_2$, for introducing no error in Step 4 and Step 5, the final state of Eve's probe should be independent of not only the operation of $P_n$ but also the measurement results of $P_n$ and $TP_2$. As a result, Eve has no access to $m_n$, let alone $p_n$.

In addition, there are other two scenarios: one is that Eve performs $U_E$ on the particle from $TP_1$ to $P_n$ and performs $U_F$ on the particle from $TP_2$ to $TP_1$; the other is that Eve applies $U_E$ on the particle from $P_n$ to $TP_2$ and applies $U_F$ on the particle from $TP_2$ to $TP_1$. After a similar proof as above, it is easy to find that in each of these two scenarios, Eve acquires no information about $m_n$ either, not to mention $p_n$.

## 4.2 Participant attacks

In 2007, Gao *et al.* [39] put forward a novel concept of attack named as participant attack for the first time. Participant attacks are generally more serious and are worth being paid more attentions to. In this section, we consider four kinds of participant attacks.

(1) The participant attack from one dishonest user

In the proposed MSQPC protocol, it is easy to see that each user is of equal importance. Without loss of generality, assume that $P_1$ is the dishonest user who tries to steal the remaining $N-1$ users' private inputs. $P_1$ may try her best to obtain the secret input of $P_j$ ( $j=2,3,\ldots,N$ ) by launch all possible attacks. In this protocol, $P_1$ is independent from $P_j$, $TP_1$ and $TP_2$. As a result, when $P_1$ launches her attacks, she essentially plays the role of an outside attacker. As illustrated in Sect.4.1, her illegal behaviors can be inevitably detected.

In addition, $P_1$ may hear of $c_j$ when $P_j$ sends it to $TP_1$ in Step 6. However, although $P_1$ knows $k_i$, she still cannot get $p_j^i$ ( $i=1,2,\ldots,L$ ) from $c_j^i$, due to lack of $m_j^i$. $P_1$ may hear the final comparison results from $TP_1$ in Step 7. However, it is still helpless for her to know $p_j^i$.

(2) The participant attack from two or more dishonest users

Here, we discuss the extreme situation that $N-1$ dishonest users collude together to obtain the private input of the remaining one user. Without loss of generality, suppose that $P_1,P_2,\ldots,P_{b-1},P_{b+1},\ldots,P_N$ try to get the private input of $P_b$, where $b=2,3,\ldots,N-1$. In this protocol, each of $P_1,P_2,\ldots,P_{b-1},P_{b+1},\ldots,P_N$ is independent from $P_b$, $TP_1$ and $TP_2$. Hence, when $P_1,P_2,\ldots,P_{b-1},P_{b+1},\ldots,P_N$ conspire to impose their attacks, they act as an outside attacker in fact. As a result, as proved in Sect.4.1, their attacks can be discovered undoubtedly.

In addition, $P_1, P_2, \ldots, P_{b-1}, P_{b+1}, \ldots, P_N$ may hear of $c_b$ while it is sent from $P_b$ to $TP_1$ in Step 6. Although $P_1, P_2, \ldots, P_{b-1}, P_{b+1}, \ldots, P_N$ knows $k_i$ ,they still cannot decode out $p_b^i$ ( $i = 1, 2, \ldots, L$ ) from $c_b^i$, because of being short of $m_b^i$. $P_1, P_2, \ldots, P_{b-1}, P_{b+1}, \ldots, P_N$ may hear the final comparison results from $TP_1$ in Step 7. Unfortunately, they still have no way to gain $p_b^i$.

(3) The participant attack from semi-honest $TP_1$

In the proposed MSQPC protocol, $TP_1$ is not allowed to conspire with anyone else. Apparently, $TP_1$ automatically knows $m_n^i$, where $n = 1, 2, \ldots, N$ and $i = 1, 2, \ldots, L$. Moreover, $TP_1$ gets $c_n$ when $P_n$ sends it to her in Step 6. Nevertheless, $TP_1$ has no idea about $k_i$ ,which means that she cannot infer $p_n^i$ from $c_n^i$. In addition, although $TP_1$ can calculate out the final comparison results in Step 7, she still cannot get $p_n^i$.

(4) The participant attack from semi-honest $TP_2$

In the proposed MSQPC protocol, $TP_2$ is not permitted to collude with anyone else. $TP_2$ naturally knows $m_n^i$, where $n = 1, 2, \ldots, N$ and $i = 1, 2, \ldots, L$. In addition, $TP_2$ may hear of $c_n$ when $P_n$ informs $TP_1$ of it in Step 6. Unfortunately, $TP_2$ still cannot decode out $p_n^i$ from $c_n^i$, due to lack of $k_i$. Furthermore, $TP_2$ may hear the final comparison results from $TP_1$ in Step 7, but still has no knowledge about $p_n^i$.

## 5 Discussions and conclusions

Ref.[32] adopts the qudit efficiency, which is converted from the qubit efficiency defined in Ref.[40], to compute the efficiency of a quantum communication protocol adaptive for the $d$ -dimensional system. According to Ref.[32], the qudit efficiency can be depicted as

$$\eta = \frac{\theta}{\sigma + \mu}, \tag{24}$$

where $\sigma$ , $\mu$ and $\theta$ are the number of qudits used, the length of classical information consumed during the classical communication and the length of private inputs compared, respectively. In the following, we calculate the qudit efficiency of the proposed MSQPC protocol by neglecting the classical resources consumed during the security check processes and the resources consumed for producing the pre-shared key sequence $K$ .

In the proposed MSQPC protocol, the length of $p_n$ ( $n = 1, 2, \ldots, N$ ) is $L$ , so we obtain $\theta = L$ . $TP_1$ needs to prepare $S_n$ , whose length is $16L$ ; after getting the qudits from $TP_1$ , when $P_n$ enters into the MEASURE mode, she needs to generate $8L$ qudits; after receiving the qudits from $P_n$ , when $TP_2$ enters into the MEASURE mode, she needs to produce $8L$ qudits; so we have $\sigma = 16L \times N + 8L \times N + 8L \times N = 32LN$ . Furthermore, $P_n$ needs to send $c_n$ to $TP_1$ , so we get $\mu = L \times N = LN$ . Therefore, the qudit efficiency of the proposed MSQPC protocol is equal to $\eta = \frac{L}{32LN + LN} = \frac{1}{33N}$ .

In addition, we compare the proposed MSQPC protocol with previous SQPC protocols in detail, and describe the comparison outcomes in Table 2. From Table 2, we easily know that the proposed MSQPC protocol exceeds the protocols of Ref.[31], Ref.[34] and Ref.[35] in quantum resources, as $d$ -dimensional single-particle states are much easier to prepare than $d$ -dimensional Bell states and $d$ -dimensional GHZ states; the proposed MSQPC protocol defeats the second protocol of Ref.[33] in usage of unitary operation, as it doesn't need any unitary operation; the proposed MSQPC protocol exceeds the protocols of Ref.[31], Ref.[34] and Ref.[35] in TP's quantum measurement, as it doesn't require $d$ -dimensional Bell state measurements or $d$ -dimensional GHZ state measurements; and the proposed MSQPC protocol is the only protocol which can judge the size relationship of private inputs from more than two classical users within one execution of protocol.

In short, in this paper, the first MSQPC protocol which can judge the size relationship of private inputs from more than two classical users within one execution of protocol is designed by using $d$ -dimensional single-particle states. This protocol has two TPs, one possessing complete quantum capabilities and the other owning limited quantum capabilities. Both TPs are permitted to misbehave on their own but cannot conspire with anyone else. This protocol requires neither quantum entanglement swapping nor unitary operations. This protocol only requires TPs to implement $d$ -dimensional single-particle measurements. This protocol can prevent both the outside attacks and the participant attacks.

Table 2 The comparison outcomes of the proposed MSQPC protocol with previous SQPC protocols

| | Ref.[31] | Ref.[32] | The first protocol of Ref.[33] | The second protocol of Ref.[33] | Ref.[34] | Ref.[35] | Our protocol |
|---|---|---|---|---|---|---|---|
| Quantum resources | $d$ -dimensional | $d$ -dimensional | $d$ -dimensional | $d$ -dimensional | $d$ -dimensional | $d$ -dimensional | $d$ -dimensional |
| | Bell states | single-particle states | single-particle states | single-particle states | Bell states | GHZ states | single-particle states |
| Number of users | 2 | 2 | 2 | 2 | 2 | 2 | $N$ |
| Number of TP | 1 | 1 | 1 | 1 | 1 | 1 | 2 |
| Type of TP | Semi-honest | Semi-honest | Semi-honest | Semi-honest | Semi-honest | Semi-honest | Semi-honest |
| Usage of pre-shared key | Yes | Yes | Yes | Yes | Yes | Yes | Yes |
| Comparison of size relationship | Yes | Yes | Yes | Yes | Yes | Yes | Yes |
| Usage of quantum entanglement swapping | No | No | No | No | No | No | No |
| Usage of unitary operation | No | No | No | Yes | No | No | No |
| TP's quantum measurement | $d$-dimensional Bell state measurements and $d$-dimensional single-particle measurements | $d$-dimensional single-particle measurements | $d$-dimensional single-particle measurements | $d$-dimensional single-particle measurements | $d$-dimensional Bell state measurements and $d$-dimensional single-particle measurements | $d$-dimensional GHZ state measurements, $d$-dimensional Bell state measurements and $d$-dimensional single-particle measuremen | $d$-dimensional single-particle measurements |
| | | | | | | ts | |
| Classical users' quantum measurement | $d$-dimensional single-particle measurements | $d$-dimensional single-particle measurements | $d$-dimensional single- particle measurements | $d$-dimensional single- particle measurements | No | $d$-dimensional single-particle measurements | $d$-dimensional single-particle measurements |
| TP's knowledge about the comparison result | Yes | No | Yes | Yes | Yes | Yes | Yes |

## Acknowledgments

Funding by the National Natural Science Foundation of China (Grant No.62071430 and No.61871347) and the Fundamental Research Funds for the Provincial Universities of Zhejiang (Grant No.JRK21002) is gratefully acknowledged.

## Reference

[1] Bennett, C.H., Brassard, G.: Quantum cryptography: public key distribution and coin tossing. In:Proceedings of the IEEE International Conference on Computers, Systems and Signal Processing,Bangalore, pp.175-179 (1984)

[2] Yao, A.C.: Protocols for secure computations. In Proc. of the 23rd Annual IEEE Symposium on Foundations of Computer Science, pp.160-164 (1982)

[3] Yang, Y.G., Wen, Q.Y.: An efficient two-party quantum private comparison protocol with decoy photons and two-photon entanglement. J. Phys. A: Math. Theor. 42(5):055305 (2009)

[4] Lin, S., Sun, Y., Liu, X.F., Yao, Z.Q.: Quantum private comparison protocol with $d$ -dimensional Bell states. Quantum Inf. Process.12: 559-568 (2013)

[5] Guo, F.Z., Gao, F., Qin, S.J., Zhang, J., Wen, Q.Y.: Quantum private comparison protocol based on entanglement swapping of $d$ -level Bell states. Quantum Inf. Process.12(8):2793-2802(2013)

[6] Luo, Q.B., Yang, G.W., She, K., Niu, W.N., Wang, Y.Q.: Multi-party quantum private comparison protocol based on $d$ -dimensional entangled states. Quantum Inf. Process. 13: 2343-2352 (2014)

[7] Ye, C.Q., Ye, T.Y.: Multi-party quantum private comparison of size relation with $d$ -level single-particle states. Quantum Inf. Process. 17(10):252 (2018)

[8] Song, X., Wen, A., Gou, R.: Multiparty quantum private comparison of size relation based on single-particle states. IEEE Access 99:1-7 (2019)

[9] Cao, H., Ma, W.P., Lü, L.D., He, Y.F., Liu, G.: Multi-party quantum comparison of size based on $d$ -level GHZ states. Quantum Inf. Process. 18: 287 (2019)

[10] Chen, F.L., Zhang, H., Chen, S.G., Cheng, W.T.: Novel two-party quantum private comparison via quantum walks on circle. Quantum Inf. Process. 20(5):1-19 (2021)

[11] Tseng, H.Y., Lin, J., Hwang, T.: New quantum private comparison protocol using EPR pairs. Quantum Inf. Process. 11:373-384 (2012)

[12] Chang, Y.J., Tsai, C.W., Hwang, T.: Multi-user private comparison protocol using GHZ class states. Quantum Inf. Process.12(2):1077-1088 (2013)

[13] Ji, Z.X., Ye, T.Y.: Quantum private comparison of equal information based on highly entangled six-qubit genuine state. Commun. Theor. Phys. 65(6):711-715 (2016)

[14] Ye, T.Y.: Multi-party quantum private comparison protocol based on entanglement swapping of Bell entangled states. Commun. Theor. Phys. 66(3):280-290 (2016)

[15] Ye, T.Y.: Quantum private comparison via cavity QED. Commun. Theor. Phys. 67(2):147-156 (2017)

[16] Ye, T.Y., Ji, Z.X.: Two-party quantum private comparison with five-qubit entangled states. Int. J. Theor. Phys. 56(5):1517-1529 (2017)

[17] Ye, T.Y., Ji, Z.X.: Multi-user quantum private comparison with scattered preparation and one-way convergent transmission of quantum states. Sci. China Phys. Mech. Astron. 60(9):090312 (2017)

[18] Ji, Z.X., Ye, T.Y.: Multi-party quantum private comparison based on the entanglement swapping of $d$ -level Cat states and $d$ -level Bell states. Quantum Inf. Process. 16(7):177 (2017)

[19] Ye, C.Q., Ye, T.Y.: Circular multi-party quantum private comparison with $n$ -level single-particle states. Int. J. Theor. Phys. 58:1282-1294 (2019)

[20] Ye, T.Y., Hu, J.L.: Multi-party quantum private comparison based on entanglement swapping of Bell entangled states within $d$ -level quantum system. Int. J. Theor. Phys. 60(4): 1471-1480 (2021)

[21] Boyer, M., Kenigsberg, D., Mor, T.: Quantum key distribution with classical Bob. Phys. Rev. Lett. 99(14):140501 (2007)

[22] Ye, T.Y., Li, H.K., Hu, J.L.: Semi-quantum key distribution with single photons in both polarization and spatial-mode degrees of freedom. Int. J. Theor. Phys. 59: 2807-2815 (2020)

[23] Ye, T.Y., Geng, M.J., Xu, T.J., Chen, Y.: Efficient semiquantum key distribution based on single photons in both polarization and spatial-mode degrees of freedom. Quantum Inf. Process. 21:123 (2022)

[24] Chou, W.H., Hwang, T., Gu, J.: Semi-quantum private comparison protocol under an almost-dishonest third party. https://arxiv.org/abs/1607.07961 (2016)

[25] Ye, T.Y., Ye. C.Q.: Measure-resend semi-quantum private comparison without entanglement. Int. J. Theor. Phys. 57(12):3819-3834 (2018)

[26] Thapliyal, K., Sharma, R.D., Pathak, A.: Orthogonal-state-based and semi-quantum protocols for quantum private comparison in noisy environment. Int. J. Quantum Inf., 16(5): 1850047 (2018)

[27] Lang, Y.F.: Semi-quantum private comparison using single photons. Int. J. Theor. Phys. 57: 3048-3055 (2018)

[28] Lin, P.H., Hwang, T., Tsai, C.W.: Efficient semi-quantum private comparison using single photons. Quantum Inf. Process.18:207 (2019)

[29] Jiang, L.Z.: Semi-quantum private comparison based on Bell states. Quantum Inf. Process., 19: 180 (2020)

[30] Ye, C.Q., Li, J., Chen, X.B. Yuan. T.: Efficient semi-quantum private comparison without using entanglement resource and pre-shared key. Quantum Inf. Process. 20:262 (2021)

[31] Zhou, N.R., Xu, Q.D., Du, N.S., Gong, L.H.: Semi-quantum private comparison protocol of size relation with $d$ -dimensional Bell states. Quantum Inf. Process. 20:124 (2021)

[32] Geng, M.J., Xu, T.J., Chen,Y., Ye, T.Y.: Semiquantum private comparison of size relationship based $d$ -level single-particle states. Sci. Sin. Phys. Mech. Astron. 52(9): 290311 (2022)

[33] Li, Y.C., Chen, Z.Y., Xu, Q.D., Gong, L.H.: Two semi-quantum private comparison protocols of size relation based on single particles. Int. J. Theor. Phys. 61: 157 (2022)

[34] Luo, Q.B., Li, X.Y., Yang, G.W., Lin, C.: A mediated semi-quantum protocol for millionaire problem based on high-dimensional Bell states. Quantum Inf. Process. 21:257 (2022)

[35] Wang, B., Liu, S.Q., Gong, L.H.: Semi-quantum private comparison protocol of size relation with $d$ -dimensional GHZ states. Chin. Phys. B 31: 010302 (2022)

[36] Krawec, W.O.: Mediated semiquantum key distribution. Phys. Rev. A 91(3):032323 (2015)

[37] Yang, Y.G., Xia, J., Jia, X., Zhang, H.: Comment on quantum private comparison protocols with a semi-honest third party. Quantum Inf. Process.12:877-885 (2013)

[38] Qin, H., Dai, Y.: Dynamic quantum secret sharing by using $d$ -dimensional GHZ state. Quantum Inf. Process. 16(3): 64 (2017)

[39] Gao, F., Qin, S.J., Wen, Q.Y., Zhu, F.C.: A simple participant attack on the Bradler-Dusek protocol. Quantum Inf. Comput.7:329 (2007)

[40] Cabello, A.: Quantum key distribution in the Holevo limit. Phys. Rev. Lett. 85:5635 (2000)